\documentclass[amsmath,amssymb,prl]{revtex4}

\newcommand{\beq}{\begin{equation}}
\newcommand{\eeq}{\end{equation}}
\newcommand{\beqa}{\begin{eqnarray}}
\newcommand{\eeqa}{\end{eqnarray}}
\newcommand{\non}{\nonumber}

\begin{document}

\title{Supersymmetric quenched complexity in the Sherrington-Kirkpatrick model}

\author{Alessia Annibale, Andrea Cavagna, Irene Giardina, Giorgio Parisi}

\affiliation{Dipartimento di Fisica, Universit\`a di Roma ``La Sapienza'' and \\
Center for Statistical Mechanics and Complexity, INFM Roma 1, \\
Piazzale Aldo Moro 2, 00185 Roma,  Italy}

\date{\today}

\begin{abstract}
By using the BRST supersymmetry we compute the quenched complexity of the TAP 
states in the SK model. We prove that the BRST complexity is equal to the Legendre 
transform of the static free energy with respect to the largest replica symmetry 
breaking point of its overlap matrix.
\end{abstract}

\maketitle

A key issue in the physics of complex systems is the computation of the entropy of
the metastable states, normally called complexity in spin-glasses, and configurational 
entropy in structural glasses and supercooled liquids. 
A knowledge of the complexity is crucial for understanding the dynamics
of a system when this is heavily influenced by strong metastability effects. 
Moreover, in some theoretical frameworks, the drop in the number of accessible 
states leads to an ergodicity breaking transition. In this context the complexity is essential 
also from the thermodynamic point of view, as in the Adam-Gibbs 
theory of the thermodynamic glass transition \cite{adam}.

Despite its enormous theoretical relevance, there are few analytic calculations
of the complexity in glassy systems, and this for a very good reason. 
In a nutshell, to find the complexity we have to compute the number of local 
minima (metastable states) of some state function, which is typically highly 
nontrivial. Just to fix ideas, we may think that this function is the Hamiltonian $H$. 
To compute the complexity, we must impose that the gradient of $H$ (the force) 
vanishes in the local minima, and we have to include as a normalization factor 
the second derivative of $H$ (the Hessian). Moreover, we may want to classify the 
metastable states according to the value of $H$, that is to their height in the landscape. 
Therefore, beside the force and the Hessian, we must include the state 
function itself in the calculation. Computing the complexity is thus a formidable 
technical task, since we have to deal with {\it three} very complicated
functions: $H$, $\partial H$, and $\partial^2 H$. In comparison, the calculation 
of the partition function, which just involves $H$, is an easy business.

This apparent difficulty in the calculation of the complexity stems from the fact that
most methods treat $H, \partial H$ and $\partial^2H$ as three independent objects, when 
of course they are not. Every calculation which fails to capture the fact that it is 
essentially just {\it one} function, $H$, that we are dealing with, effectively wastes 
a crucial information. It would be therefore important to find a tool which exploits 
this information to simplify the calculation of the complexity.
The Becchi-Rouet-Stora-Tyutin (BRST) supersymmetry  \cite{brs,tito} seems to be such a tool. 
As it was first noted in \cite{juanpe} for a particular model, a BRST
calculation of the complexity is in fact equivalent to the one of the partition function. 
This is indeed what we expect from a method which does {\it not} treat $H, \partial H$ and 
$\partial^2H$ as independent functions. 

The formal equivalence between 
complexity and standard thermodynamics found in \cite{juanpe} by means of the BRST 
supersymmetry is a very important theoretical issue.
In the context of spin-glasses the existence of such a connection 
has been much investigated in the past \cite{bm1,bm2,bm3,ddy,bmy,monasson,
franzparisi,potters,crisatap}.
In a classic paper \cite{bm1}, Bray and Moore first calculated the complexity of the 
Sherrington-Kirkpatrick (SK) model \cite{sk} by counting the number of local minima of the 
Thouless-Anderson-Palmer (TAP) free energy \cite{tap}, 
which in mean-field spin-glasses is the state function discussed above.
The same authors also noted in \cite{bm2} some deep formal connections 
between TAP complexity and standard thermodynamics, while De Dominicis and 
Young showed in \cite{ddy} that TAP and static approaches were 
in fact equivalent, once some key hypothesis were made. These 
studies culminated in a remarkable work \cite{bmy}, where Bray, Moore and Young 
uncovered a sort of Legendre transform relationship between TAP complexity 
and static free energy.

A method to compute the complexity which does not rely on the existence of 
a TAP free energy, was introduced by Monasson \cite{monasson}, and by Franz and Parisi 
\cite{franzparisi}. The basic idea is to introduce a coupling between 
different systems forcing them to live in the same metastable state. The free 
energy cost of such a constrained super-system is equal 
to the entropic contribution of the metastable states, that is the complexity.
Within this approach, close connections between complexity and thermodynamics, 
similar to those found in the TAP context in \cite{bmy}, were found. In particular, 
in spin-glass models with one step of replica 
symmetry breaking \cite{1rsb,mpv}, the formulation of Monasson shows that 
the complexity is equal to the Legendre transform of the static free energy with respect 
to the breaking point $x$ of the overlap matrix \cite{monasson,crisatap}. 

Despite all these investigations, it is fair to say that a general formal 
connection between complexity of the metastable states and static free energy, has 
not been proved yet. In particular, it is unclear how the Legendre transform method 
of \cite{monasson} should be used in systems with more than one step of replica symmetry 
breaking, as the SK model. In fact, none of the previous SK investigations \cite{bm2,bmy} 
succeeded in proving the existence of a sharp Legendre transform relationship as in one-RSB 
systems.

In this Letter we find for the first time an exact connection between complexity of 
the metastable states and static free energy in the SK model: 
we prove that the quenched TAP complexity obtained  by means of the BRST supersymmetry
is the Legendre transform of the static free energy with respect to the {\it largest} 
breaking point of its overlap matrix. Our result confirms the validity of the Legendre 
transform method of \cite{monasson,franzparisi,potters}, and its consistency with the
investigations of \cite{bm2,bm3,ddy,bmy}. 
Moreover, our findings strongly suggest that the BRST supersymmetry should 
be considered as an essential tool for the computation of the complexity 
in more general glassy systems.

The complexity of the TAP states with free energy density $f$, at inverse temperature $\beta$,
is defined as \cite{bm1},
\beq
\Sigma(\beta,f) 
= \frac{1}{N}\log \sum_{\alpha=1}^{\cal N} \delta\left[ N\beta f - \beta F_{TAP}(m^\alpha)\right] 
= \frac{1}{N}\log \int dr\ e^{N r \beta f}\  \sum_{\alpha=1}^{\cal N} e^{-r\beta F_{TAP}(m^\alpha)} \ ,
\label{riz}
\eeq
where $m^\alpha\equiv\{m_i^\alpha\}$, are the local magnetizations at site $i=1\dots N$, 
in state $\alpha=1\dots\cal N$. A state $m^\alpha$ is defined as a local minimum of 
the TAP free energy $F_{TAP}(m)$ \cite{tap}. If we define the thermodynamic potential 
$\Psi(\beta,r)$,
\beq
\exp\left( -\beta N r \, \Psi \right)
\equiv
\sum_{\alpha=1}^{\cal N} e^{-r\beta  F_{TAP}(m^\alpha)} \ ,
\label{psidef}
\eeq
we can use the steepest descent method in (\ref{riz}), and obtain the complexity as the Legendre transform
of $\Psi(\beta,r)$,
\beq
\Sigma(\beta,f) = \beta r f -\beta r\,\Psi(\beta,r) \ , 
\label{heidi}
\eeq
where the parameter $r=r(\beta,f)$ is fixed by the equation,
\beq
\Psi(\beta,r) + r \,\frac{\partial\Psi(\beta,r)}{\partial r}=f \ .
\label{nonno}
\eeq
From (\ref{psidef}) we see that for $r=1$ 
the potential $\Psi$ must be equal to the standard static free energy of the system
$F(\beta)$, calculated in the TAP context. This calculation was first performed
in \cite{ddy}, where it was shown that the relation $\Psi(\beta,r=1)=F(\beta)$ only 
held if some suitable assumptions were made. In \cite{brst1} it was proved that the 
assumptions used in \cite{ddy} were in fact a general consequence of the BRST 
supersymmetry. However, the situation was less clear for generic values of $r$, 
since the calculations of \cite{bm3,bmy} for $r\neq 1$ seemed to explicitly break 
the BRST invariance \cite{nota}.
In what follows we perform a supersymmetric quenched calculation of $\Psi(\beta,r)$, 
and prove that this potential is intimately related to the static free 
energy $F(\beta)$ even for $r\neq 1$. The TAP free energy for the SK model is given by 
\cite{tap},
\beq
\beta F_{TAP}(m)=-\frac{\beta}{2} \sum_{ij} J_{ij} m_i m_j 
+ \frac{1}{\beta} \sum_i \phi_0(m_i) \ ,
\eeq
with,
\beq
\phi_0(m) =
\frac{1}{2}\log(1-m^2) + m\,\tanh^{-1}(m) -\log 2 -\frac{\beta^2}{4}(1-q)^2 \ .
\label{phi0}
\eeq
The variable $q$ is  the self-overlap of the TAP states,
$q=\frac{1}{N}\sum_i m_i^2$. The quenched couplings $J$ are random variables with Gaussian 
distribution and variance $N$. From (\ref{psidef}) we have that the {\it quenched} potential 
$\Psi(\beta,r)$ is,
\beq
- \beta r\, \Psi(\beta,r)
= \frac{1}{nN}\log\overline{\rho(\beta,r|J)^n} \ ,
\label{nutria}
\eeq
with,
\beq
\rho(\beta,r|J)
=\sum_{\alpha=1}^{\cal N} e^{-r\beta  F_{TAP}(m^\alpha)} 
=\int \prod_i dm_i\ \delta(\partial_i F_{TAP}(m)) \
 |\det (\partial_i \partial_j F_{TAP}(m))| 
\ e^{-\beta r\,F_{TAP}(m)} \ .
\label{rho}
\eeq
In (\ref{nutria}) we have $N\to\infty$ and $n\to 0$, and the over-bar indicates an average 
over the disorder. 
As usual, the modulus of the determinant will be dropped. 
This amounts to assume that at sufficiently low temperatures
the largest part of TAP solutions are minima \cite{brst1}. Of course, any method which
drops the modulus is doomed to fail if stable minima are subdominant with respect to 
unstable saddles.  
After introducing the commuting fields $x_i$ to implement the delta functions,
and the anti-commuting (Grassmann) fields $\bar\psi_i,\psi_i $ for the determinant, 
we find,
\beq
\rho(\beta,r|J)
= \int {\cal D}m\, {\cal D}x\,{\cal D}\bar\psi\, {\cal D}\psi \ \
e^{\beta S(m,x,\bar\psi,\psi)} \ ,
\label{pino}
\eeq
where the action $S$ is given by,
\beq
S(m,x,\bar\psi,\psi)= \sum_i x_i \partial_i F_{TAP}(m) + \sum_{ij}\bar\psi_i  \psi_j  
\partial_i \partial_j F_{TAP}(m) - r F_{TAP}(m) \ .
\label{action}
\eeq
By averaging $\rho(\beta,r|J)^n$  over the disorder we obtain 
the following effective action,
\beqa
\beta S &=& \frac{\beta^2}{2 N} \left[\sum_{ab}^n \left(\sum_i^N x_i^a x_i^b\right) 
\left(\sum_j^N m_j^a m_j^b\right) +\sum_{ab}^n \left(\sum_i^N x_i^a m_i^b\right)^2 - 
\sum_{ab}^n \left(\sum_i^N \bar\psi_i^a \psi_i^b\right)^2\right] \
\non
\\
&+&  \frac{\beta^2}{2 N} \left[\frac{r^2}{2} \sum_{ab}^n \left(\sum_i^N m_i^a m_i^b\right)^2 -
2 r \sum_{ab}^n \left(\sum_i^N m_i^a x_i^b\right) \left(\sum_j^N m_j^a m_j^b\right)\right] 
\non
\\
&+& \sum_a^n \sum_i^N \left[x_i^a \phi_1(m_i^a) + 
\bar\psi_i^a \psi_i^b \phi_2(m_i^a) - r  \phi_0(m_i^a)\right] \ ,
\eeqa
where $\phi_1=\partial_m\phi_0$ and $\phi_2=\partial_m^2\phi_0$. 
The scalar overlap $q$ will now be generalized by introducing the overlap matrix
$q_{ab}=m^a\cdot m^b$. This form of the action is different from the one in \cite{bm3,bmy}, 
where the delta 
function enforcing the TAP equations was used to eliminate the factor
$J_{ij}m_i^a m_j^a$ in (\ref{action}). This is a crucial point: this method 
explicitly {\it breaks} the BRST invariance of the action, which is the crucial 
tool to establish the exact connection with the static free energy. We therefore 
do not use this method, and keep the whole BRST invariant action \cite{nota}.

In order to linearize the quadratic terms we introduce the usual Lagrange multipliers, 
$m^am^b\to \lambda_{ab}$, $m^a x^b\to w_{ab}$, $\bar\psi^a\psi^b
\to t_{ab}$ \cite{brst1}. 
After this is done, the integrals in $x$ and $\bar\psi,\psi$ 
become Gaussian and can be performed explicitly. Moreover, the action factorizes and
for $N\to\infty$ we can use the steepest descent method, to get,
\beq
- \beta r\,\Psi(\beta,r) = \lim_{n\to 0} \, \frac{1}{n}\, \left[\, \Sigma_0 +
\log\int\prod_a dm^a e^{{\cal L}(m^a)}\, \right] \ .
\eeq
Following \cite{bm1,bm3,brst1} we define,
\beqa
B_{ab}&\equiv&\beta^2(1-q_{aa})\, \delta_{ab} + t^{ab} 
\\
\Delta_{ab}&\equiv& -\beta^2(1-q_{aa})\, \delta_{ab} - w^{ab} \ ,
\eeqa
and therefore obtain (details will be given elsewhere), 
\beqa
\Sigma_0 = \
\frac{1}{2\beta^2}\sum_{ab} (B_{ab}^2-\Delta_{ab}^2) - 
\sum_a\left(B_{aa}+\Delta_{aa}\right) (1 - q_{aa}) 
\phantom{praaaaaaaaaaaaaaaaaaaaaaaaaaaaaaaat}
\non
\\
\phantom{praaaaaaaaaaaaaaaaaaaaat}
- \sum_{ab}\left[
\frac{\beta^2}{4}r^2 \,q_{ab}^2 + \lambda^{ab} q_{ab} - r  \Delta^{ab} q_{ab} \right]
- \frac{1}{2} \log[(2\pi\beta^2)^n \det q_{ab}] 
\label{sigma} 
\\
{\cal L}(m^a)= - r \sum_a \phi_0(q_{aa},m^a) 
+ \sum_{ab} \lambda^{ab} m^a m^b + \log\det\left(\frac{\delta_{ab}}{1-m_a^2}+B_{ab}\right) 
\phantom{praaaaaaaaaaaaaaaaaaaaat}
\non
\\
\phantom{praaaaaaaaaaaaaaaaaaaaat}
- \frac{1}{2\beta^2}\sum_{ab} 
\left[\tanh^{-1}m^a - \sum_c \Delta^{ac} m^c\right] 
q_{ab}^{-1} 
\left[\tanh^{-1}m^b - \sum_c \Delta^{bc} m^c\right] \ .
\label{elle}
\eeqa
The parameters $\Delta_{ab}, B_{ab}, \lambda_{ab}$ and $q_{ab}$ must be fixed by
the saddle point equations, and it is easy to show that $B_{ab}=0$ is solution.
It is important at this point to consider the role of the supersymmetry.
In \cite{juanpe} it was noted that action (\ref{action}) is invariant under a generalization 
of the BRST supersymmetry \cite{brs,tito}, 
\[
\delta m_i = \epsilon\, \psi_i \quad\quad
\delta x_i = \epsilon \, r \, \psi_i \quad\quad
\delta \bar\psi_i = -\epsilon\, x_i \quad\quad
\delta \psi_i = 0 
\ .
\]
where  $\epsilon$ is an infinitesimal Grassmann parameter.
If we calculate the variation of $m_i \bar\psi_i$ and  $x_i\bar\psi_i$ \cite{juanpe,brst1}, 
we obtain the two BRST equations,
\beqa
\langle \bar\psi_i \psi_i \rangle &+& \langle m_i x_i \rangle =0 
\label{brst1}
\\
r \,\langle \bar\psi_i \psi_i \rangle &+& \langle x_i x_i \rangle =0  \ \ .
\label{brst2}
\eeqa
After some algebra, these equations become,
\beqa
\Delta_{ab}&=& \beta^2 q_{ab}\, r 
\label{brst11}
\\
\lambda_{ab}&=& \frac{1}{2}\, \beta^2 r^2 \, q_{ab} \ ,
\label{brst22}
\eeqa
and it is possible to show that the remaining saddle point equations are indeed satisfied by 
(\ref{brst11}) and (\ref{brst22}). The only saddle point equation we are left with is
obtained by doing the variations of (\ref{sigma}) and (\ref{elle}) with respect to 
$\lambda_{ab}$. This gives,
\beq
q_{ab} =\langle\langle m^a m^b \rangle\rangle \ ,
\label{sad1} \
\eeq
where the average is performed with the distribution $\exp({\cal L}(m^a))$.
If we use equations  (\ref{brst11}), (\ref{brst22}) and (\ref{sad1})
into (\ref{sigma}) and (\ref{elle}), and make the change of variable 
$m^a\to h^a=\tanh^{-1}(m^a)$, we finally obtain,
\beq
\beta\Psi(\beta,r) = 
-\log 2  
+ \frac{\beta^2}{4n}\left[r\sum_{ab}^n q_{ab}^2 - \sum_a^n  (1-q_{aa})^2\right] 
-\frac{1}{nr} \log\int\prod_a^n  
\frac{dh^a}{\sqrt{2\pi \beta^2 \det q_{ab}}}\ \, 
\cosh(h^a)^r \
e^{-\frac{1}{2\beta^2}\sum_{ab}^n h^a q_{ab}^{-1} h^b}
 \ .
\label{generale}
\eeq
Expression (\ref{generale}) is different from the one of \cite{bm3,bmy}: the BRST 
supersymmetry automatically selects one saddle point in the space of parameters,
and in so doing it drastically reduces the number of parameters, compared to \cite{bm3,bmy}.
The computation of $\Psi(\beta,r)$ has at this point the same
degree of difficulty as the one of the standard free energy $F(\beta)$, with just one
overlap matrix $q_{ab}$ to be fixed variationally. We shall now show that the 
connections between $\Psi(\beta,r)$ and $F(\beta)$ are in 
fact much deeper than that.

The general form of the quenched free energy in the SK model is \cite{sk},
\beq
\beta F(\beta)= - \frac{\beta^2}{4}+ \frac{\beta^2}{2n_s}\sum_{\alpha>\beta}^{n_s} Q_{\alpha\beta}^2 
- \frac{1}{n_s}\log\sum_{[\sigma^\alpha]}\exp\left[\frac{\beta^2}{2}
\sum_{\alpha\neq \beta}^{n_s} Q_{\alpha\beta}\, \sigma^\alpha \sigma^\beta \right]\ ,
\label{statica}
\eeq
where $Q_{\alpha\beta}$ is the $n_s\times n_s$ overlap matrix, with $n_s\to 0$. 
If in (\ref{generale}) we use the relation,
\beq
\cosh(h_a)^{r} = \frac{1}{2^r}\, \sum_{[\tau_a^\mu=\pm 1]}e^{h_a\, \sum_\mu^r \tau_a^\mu} \ ,
\eeq
we obtain,
\beq
\beta\Psi(\beta,r)
= -\frac{\beta^2}{4}  
+\frac{\beta^2}{2n}\left[ 
r \sum_{a>b}^n q_{ab}^2
+ \frac{r-1}{2}\sum_a^n q_{aa}^2 \right] 
-\frac{1}{nr} \log\sum_{[\tau_a^\mu]}
\exp\left[\frac{\beta^2}{2}\left(
\sum_{ab}^n \sum_{\mu \nu}^r \tau_a^\mu q_{ab} \tau_b^\nu
- \sum_a^n\sum_\mu^r q_{aa}\right) \right] \ .
\label{uruk}
\eeq
The trace terms in equations (\ref{statica})
and (\ref{uruk}) suggest the relation $n_s= r \cdot n$. Once this identification is done, 
we can connect the $\sigma^\alpha$ spin variables ($\alpha=1,\dots, n_s$), to the $\tau_a^\mu$ 
spin variables ($a=1,\dots,n;\  \mu=1,\dots, r$) in the following way,
\[
\left(\sigma_1,\dots ,\sigma_{n_s}\right)
= \left(\tau_1^1,\dots ,\tau_1^r,
\ \dots \dots  , \tau_n^1,\dots ,\tau_n^r\right) \ .
\]
Let us now assume that the potential $\Psi(\beta,r)$ (and thus the TAP complexity) is 
calculated at $k$ levels of replica symmetry breaking (RSB) \cite{mpv}. The TAP overlap matrix $q_{ab}$ 
is then given by,
\beq
q_{ab}^{(k)} =q_0 +  \sum_{i=1}^{k+1}(q_i-q_{i-1}) \ \varepsilon_{ab}^{(n, y_{i})}  \ , 
\eeq
with $y_{k+1}=1$ and $\varepsilon_{ab}^{(n,1)}=\delta_{ab}$. The matrices $\varepsilon^{(n, y_{i})}$ 
are $n\times n$ ultrametric block matrices, equal to one on the diagonal blocks of size $y_i$, 
and zero elsewhere. The variables $y_i$ are thus the replica symmetry breaking points. 
In the TAP approach the diagonal of the overlap matrix, $q_{aa}= q_{k+1}$, contains  
the self-overlap of the states, and for this reason $y_{k+1}=1$. 
There are $k+1$ values of the overlap, but only $k$ nontrivial breaking points, 
and thus $q_{ab}$ is a $k$-RSB matrix. Given this form of $q_{ab}$, it is possible 
to prove (details will be given elsewhere) that,
\beq
\sum_{ab}^n\sum_{\mu\nu}^r \tau_a^\mu q_{ab}^{(k)} \tau_b^\nu - \sum_a^n\sum_\mu^r q_{aa}^{(k)}
\ = \ 
\sum_{\alpha\neq \beta}^{rn} Q_{\alpha\beta}^{(k+1)}\, \sigma_\alpha \sigma_\beta\, \ ,
\label{stella}
\eeq
where $Q_{\alpha\beta}^{(k+1)}$ is a standard $rn\times rn$ RSB matrix, with $k+1$ levels of 
replica symmetry breaking. More precisely, 
\beq
Q^{(k+1)}_{\alpha\beta}
= q_0 + \sum_{i=1}^{k+1}(q_i-q_{i-1}) \ \varepsilon_{\alpha\beta}^{(rn, r y_{i})} \ .
\eeq
From this formula we see that the entries of $Q_{\alpha\beta}^{(k+1)}$ are the same as $q_{ab}^{(k)}$, 
whereas the $k+1$ replica symmetry breaking points $x_i$ of $Q_{\alpha\beta}^{(k+1)}$ 
are rescaled by a factor $r$, that is $x_i=r\, y_i$. In particular, the {\it largest} breaking 
point $x\equiv x_{k+1}$ of the static matrix $Q_{\alpha\beta}^{(k+1)}$ is given by,
\beq
x = r \ .
\eeq
By inserting relation (\ref{stella}) into (\ref{uruk}), we finally obtain,
\beq
\Psi(\beta,r \,| q_{ab}^{(k)}) = F(\beta|\, Q_{ab}^{(k+1)}) \ ,
\label{snapora}
\eeq
We have thus proved that the thermodynamic potential $\Psi(\beta,r)$ 
calculated at the $k$ RSB level is equal to static free energy $F(\beta)$ calculated at 
the $k+1$ RSB level. The replica symmetry breaking points of the static matrix 
$Q_{\alpha\beta}$ are simply the ones of the TAP matrix $q_{ab}$ 
rescaled by the parameter $r$, and the extra $k+1$-th breaking point 
of $Q_{\alpha\beta}$ is equal to $r$. This rescaling was first noted in \cite{bmy}, 
and later in \cite{potters}, although the lack of BRST symmetry of those calculations 
prevented to prove equation (\ref{snapora}).

From equation (\ref{heidi}), and given the relation between $\Psi(\beta,r)$
and $F(\beta)$, we finally have the general Legendre equation connecting 
the quenched complexity of the TAP states to the standard 
static free energy in the SK model,
\beq
\Sigma(\beta,f) = \beta x f - \beta x F(\beta; x) \ ,
\label{tuka}
\eeq
with the largest breaking point $x$ fixed by the equation,
\beq
f = F(\beta; x) + x\, \frac{\partial F(\beta; x)}{\partial x} \ .
\label{stuka}
\eeq
This result can be summarized as follows: {\it the supersymmetric quenched complexity of the 
TAP states is
the Legendre transform of the static free energy with respect to the largest breaking point 
$x$ of its overlap matrix}. 
This result allows to compute the exact quenched complexity in the SK model, that is 
the complexity at the {\it full}- RSB level, as the Legendre transform of the full-RSB 
static free energy \cite{leuzzi}. 

It has been conjectured in \cite{riko} that in systems with more than one step 
of replica symmetry breaking the complexity of {\it clusters} at level $i$ is
given by the Legendre transform of the free energy with respect to the breaking point $x_i$. 
For $x_i=x_{max}$ clusters are just  states, and our result is recovered.
It would be interesting to study whether the conjecture of \cite{riko} can
be exactly proved within the supersymmetric formalism used here.

We thank A. Crisanti, L. Leuzzi, R. Monasson, A. Montanari, F. Ricci-Tersenghi, and T. Rizzo for 
some interesting discussions.


\begin{thebibliography}{22}

\bibitem{adam} G. Adam and J.H. Gibbs, J. Chem. Phys. {\bf 43}, 139 (1965).
\bibitem{brs} Becchi C, Rouet A and Stora R 1974 {\it Phys. Rev. Lett.} {\bf 52B} 344.
\bibitem{tito} Tyutin I V 1975 {\it Lebedev preprint} FIAN 39, unpublished. 
\bibitem{juanpe} Cavagna A, Garrahan J P and Giardina I 1998 {\it J. Phys. A: Math. Gen.} {\bf 32} 711.
\bibitem{bm1} Bray A J and Moore M A 1980 {\it J. Phys. C: Solid. St. Phys} {\bf 13} L469.
\bibitem{bm2} Bray A J and Moore M A 1980 {\it J. Phys. C: Solid. St. Phys} {\bf 13} L907.
\bibitem{bm3} Bray A J and Moore M A 1980 {\it J. Phys. A: Math. Gen.} {\bf 14} L377.
\bibitem{ddy} De Dominicis C and Young A P, 1983 {\it J. Phys. A: Math. Gen.} {\bf 16} 2063.
\bibitem{bmy} Bray A J, Moore M A and Young A P, 1984 {\it J. Phys. C: Solid St. Phys} {\bf 17} L155.
\bibitem{monasson} Monasson R 1995 {\it Phys. Rev. Lett.} {\bf 75} 2847.
\bibitem{franzparisi} Franz S. and Parisi G. 1995 {\it J. Physique I} {\bf 3} 1819.
\bibitem{potters} Potters M and Parisi G 1995 {\it Europhys. Lett.} {\bf 32} 13.
\bibitem{crisatap} Crisanti A and Sommers H-J 1995 {\it J. Phys. I France} {\bf 5} 805
\bibitem{sk} Sherrington D and Kirkpatrick S 1975 {\it Phys. Rev. Lett.} {\bf 32} 1792.
\bibitem{tap} Thouless D J, Anderson P W and Palmer R G 1977 {\it Phil. Mag.} {\bf 35} 593.
\bibitem{1rsb} Parisi G 1979 {\it Phys. Lett.} {\bf 73A} 203.
\bibitem{mpv} M\'ezard M,  Parisi G and Virasoro M.A., {\it Spin glass theory and beyond}, 
World Scientific (1987).
\bibitem{brst1} Cavagna A, Giardina I, Parisi G and Mezard M 2003 {\it J. Phys. A: Math. Gen.} 
{\bf 36} 1175.
\bibitem{nota} Very recent results, however, seem to show that the supersymmetry can effectively 
be restored after a suitable reparametrization \cite{leuzzi}.
\bibitem{leuzzi} A. Crisanti, L. Leuzzi, T. Rizzo, in preparation (2003).
\bibitem{riko} A. Montanari, F. Ricci-Tersenghi, preprint cond-mat/0301591 (2003).


\end{thebibliography}
\end{document}